\documentclass[acmsmall,nonacm]{acmart}

\newcommand{\add}[1]{\textcolor{black}{#1}}
\newcommand{\final}[1]{\textcolor{black}{#1}}

\AtBeginDocument{%
  \providecommand\BibTeX{{%
    \normalfont B\kern-0.5em{\scshape i\kern-0.25em b}\kern-0.8em\TeX}}}

\begin{document}

\title{The Phase Model of Misinformation Interventions}

\author{Hendrik Heuer}
\email{hendrik.heuer@cais-research.de}
\orcid{0000-0003-1919-9016}
\affiliation{%
  \institution{Center for Advanced Internet Studies \& University of Wuppertal}
  \city{Bochum}
  \country{Germany}}

\begin{abstract}
Misinformation is a challenging problem. This paper provides the first systematic interdisciplinary investigation of technical and non-technical interventions against misinformation. It combines interviews and a survey to understand which interventions are accepted across academic disciplines and approved by misinformation experts. Four interventions are supported by more than two in three misinformation experts: promoting media literacy, education in schools and universities, finding information about claims, and finding sources for claims. The most controversial intervention is deleting misinformation. We discuss the potentials and risks of all interventions. \final{Education-based interventions are perceived as the most helpful by misinformation experts. Interventions focused on providing evidence are also widely perceived as helpful. We discuss them as scalable and always available interventions that empower users to independently identify misinformation. We also introduce the Phase Model of Misinformation Interventions that helps} practitioners make informed \final{decisions} about which interventions to focus \final{on and how to best combine interventions}.
\end{abstract}



\maketitle

\section{Introduction}

Misinformation and related phenomena have a long history. They have been discussed in modern academic work as early as during the Second World War~\cite{allport1947psychology} and as recently as the 2023 Israel–Hamas war~\cite{nytimes_gaza_hospital}, the COVID-19 pandemic~\cite{brennen2020types}, and the 2020 United States Capitol attack~\cite{enwiki:1032713619}. In addition to this long history, there is a broad consensus that misinformation and related phenomena are a critical problem. A 2019 survey by \add{the} Pew Research Center showed that false information is an essential issue for many citizens and that many citizens frequently fact-check information. Nine of \final{ten} U.S. citizens believe that made-up news \final{cause} a great deal of confusion about the basic facts of current events~\cite{Mitchell_Pew_2019}. The survey also found that many U.S.~citizens self-report checking the facts of news stories. Whether people check the facts is influenced by their political awareness and political knowledge. Almost nine out of ten highly politically aware U.S.~citizens state that they check facts (88\%). Of those less politically aware, almost seven out of ten check facts (68\%).

Even though \add{misinformation is a significant and widely recognized problem}, CSCW scholars lack a clear direction on what \final{interventions} to prioritize and pursue. \add{This paper is the first to provide an empirically grounded overview of possible interventions against misinformation from a broad interdisciplinary perspective}. This overview can help researchers, practitioners, and decision-makers prioritize what interventions to pursue. \add{Our investigation shows which interventions are supported across disciplines. We also assess how much support the interventions receive from misinformation experts.} For this decision, it is important to understand the \add{risks} associated with different interventions. It is also \add{essential} to understand in \final{what contexts} the interventions can be applied.

\final{This paper answers the following three research questions:}

\begin{itemize}
\item \final{\textbf{RQ1:} What interventions against misinformation does an interdisciplinary group of people propose in interviews?}
\item \final{\textbf{RQ2:} What are \add{the} potential risks of deploying these interventions?}
\item \final{\textbf{RQ3:} How do misinformation experts perceive the interventions proposed by non-experts?}
\end{itemize}

To identify interventions and to understand their risk and applicability, we performed a mixed-methods investigation that combined interviews and a survey. In the interviews, a diverse group from different research fields proposed possible interventions against misinformation and discussed their risks and applicability. In a subsequent survey, misinformation experts rated whether they agreed or disagreed that certain interventions are helpful. The interviews maximized the \add{breadth} of ideas, while the survey allowed us to \add{understand} what established experts consider helpful.

Our investigation identified 11 interventions against misinformation. Eight of these interventions are supported by a majority of misinformation experts. In an analysis \add{step}, we organized the intervention proposals into four categories: educating people, labeling misinformation, providing evidence, and deleting misinformation. We also discuss the risks of each intervention. Based on our analysis, we integrate the intervention proposals into \add{the first publicly available research overview of empirically grounded interventions: the \textit{Phase Model of Misinformation Interventions}. \final{This model} can inform} the design and development of socio-technical interventions that help individuals and \final{society} as a whole to become less vulnerable to misinformation. 

In summary, we make the following contributions:

\begin{itemize}
    \item \add{We provide the first interdisciplinary overview of interventions against misinformation that are approved by misinformation experts}.
    \item \add{We contribute the first interdisciplinary discussion of risks associated with interventions against misinformation}.
    \item We assess \final{how} established misinformation experts view the intervention proposals.
    \item We introduce the Phase Model of Misinformation Interventions to help practitioners make informed decisions about what intervention to pursue.
\end{itemize}

\section{Background}

The study of misinformation and related phenomena has a long history~\cite{allport1947psychology}. We start the background section by explaining \add{the operationalization of misinformation we adopt}. Traditionally, Wardle and Derakhshan distinguish between misinformation as false information, including false connections and misleading content, and disinformation as a combination of an intent to harm and information that is false~\cite{wardle2018thinking}. For Wardle and Derakhshan, disinformation is content put in the wrong context and imposter, manipulated, or fabricated content. Informed by a recent survey conducted by Altay et al.~\cite{altay2023survey}, we do not make such a strong distinction between misinformation and disinformation. Altay et al. surveyed 150 researchers specializing in misinformation. This survey identified significant variations in the terminology employed by experts, especially in relation to the concept of intention. While 43\% of experts defined misinformation as ``false and misleading information,'' nearly one in three (30\%) described misinformation as ``false and misleading information disseminated unintentionally.'' The survey also revealed that experts with a qualitative background were more inclined to incorporate the notion of unintentionality into their definition of misinformation \add{than} their quantitative counterparts. This disparity in terminology informed us to include both false and misleading information that is shared intentionally or unintentionally. We will refer to misinformation throughout the paper.

To understand users' susceptibility to misinformation, it is fundamental to understand how the veracity of information and claims is evaluated. From a psychological perspective, Lewandowsky et al.~cite four questions that people are ``likely to attend to'' when thoughtfully evaluating the truth value of information~\cite{doi:10.1177/1529100612451018}. These include assessing (1)~whether the information is compatible with other things a user believes in, (2)~whether the information is internally coherent, (3)~whether it comes from a credible source, and (4)~whether other people believe in it. They highlight that checking a piece of information against other knowledge is effortful and requires motivation and cognitive resources. 

Jahanbakhsh et al. offer an HCI perspective on the factors contributing to users' acceptance of individual claims~\cite{farnaz2021lightweight}. Drawing from a survey involving over 1,800 individuals in the U.S., they have devised a classification system outlining the rationales behind believing a specific assertion. These rationales encompass personal firsthand knowledge, trust in \final{sources}, corroboration by other reputable sources, the presence of supporting evidence in the news story, and alignment with an individual's own experiences and observations. Conversely, skepticism toward a news story is influenced by firsthand knowledge, inconsistencies between the claim and trusted sources, \add{and} disparities with personal experiences and observations. Additionally, individuals may \add{consider the story's presentation} and assess whether the claim appears motivated by factors that could compromise honesty or introduce bias. These findings are pertinent to previous research on how users' belief in misinformation impacts their perception of a source's credibility~\cite{10.1145/3415211,swire2017processing,metzger2010social,berinsky2017rumors}.

\subsection{Interventions Against Mis- and Disinformation}

The primary contribution of this paper is exploring \final{interventions} against misinformation from an interdisciplinary perspective and providing an actionable research roadmap that is empirically grounded. \add{Our work is motivated by the insight that interventions against misinformation are rarely subject to interdisciplinary discussions, especially not at a formative stage before they are implemented. Even though numerous publications focus on interventions against misinformation, an interdisciplinary account of what interventions to develop needs to be included, especially an account that provides an in-depth qualitative perspective. Throughout the paper, we link these proposals to proposals that have already been published. The following sections provide a non-exhaustive overview of the technical and non-technical interventions} \final{proposed in prior work}.

\subsubsection{Technical Interventions}

A large body of research on technical interventions relies on machine learning and artificial intelligence~\cite{heuer2021machine,jarke2024algorithmic}. \final{Interventions focused on those who spread the misinformation~\cite{davis2016botornot,10.1145/3159652.3159677,10.1145/2872518.2890098} include Wu and Liu's TraceMiner. Their} system can classify propagation pathways of a message~\cite{10.1145/3159652.3159677}. Another example is Davis et al.'s BotOrNot\add{,} which attempts to detect social bots~\cite{davis2016botornot}. A similar goal is pursued by Hoaxy, a platform introduced by Shao et al.~\cite{10.1145/2872518.2890098}. Hoaxy monitors and visualizes how misinformation and fact-checks are spread on social media platforms like Twitter.

\final{Interventions that automatically detect misinformation~\cite{10.1145/3137597.3137600}  commonly frame the task as predicting whether a news article is a misinformation piece or not~\cite{perez-rosas-etal-2018-automatic,wang-2017-liar}}. For the misinformation detection task, Zhou et al.~distinguish between lexicon-, syntax-, semantic-, and discourse-level approaches~\cite{10.1145/3377478}. Other technical approaches found in related work take linguistic and stylistic signals and the affect of a message into account~\cite{10.1145/3274351,schuster2020limitations,Khalid_Srinivasan_2020}. Jiang and Wilson, for instance, show that linguistic signals in user comments vary significantly with the veracity of posts. They demonstrate how this can be used to detect misinformation~\cite{10.1145/3274351}. \final{Despite their potential, such approaches are frequently criticized for only capturing correlations and for not identifying causal mechanisms. This motivated researchers} in fields like critical data studies \final{to} highlight the risks of relying on correlations and the \add{vital} role training data plays in ML-based systems~\cite{d2020data,joler_nooscope_2020}. Asr et al., for instance, investigate what datasets are available to train ML-based systems to automatically detect whether a news article is reliable or not. They find that such datasets are limited and that ensuring data quality is a challenging problem~\cite{doi:10.1177/2053951719843310}. More generally, the potential of ML-based systems to make biased decisions that discriminate against specific groups or individuals is widely recognized~\cite{d2020data,eubanks2018automating,noble2018algorithms}. \final{This overview of prior work motivated us to examine misinformation interventions that go beyond examining who spreads misinformation and predicting whether a news article is a misinformation piece or not. This inspired us to systematically examine the strengths and weaknesses of all possible misinformation interventions by collecting interdisciplinary proposals.}

\subsubsection{Socio-Technical Interventions}
In addition to technical interventions, several socio-technical interventions have been examined. \add{We operationalize the term socio-technical intervention as any intervention where social and technical aspects are relevant for a solution to work.} Such socio-technical interventions span a range of approaches, including, but not limited to, trust ratings~\cite{pennycook2019crowdsourcing}, nudges~\cite{bhuiyan2018feedreflect}, collaborative fact-checking~\cite{10.1145/3308560.3316734}, and crowdsourcing~\cite{10.1145/3359209}. \add{In all \final{these} examples, both the technical implementation and the social participation of users are necessary for a solution to work. This insight relates to Brunes at al., who} studied the effect of \add{prebunking} and debunking interventions in four European countries~\cite{bruns_prebunks_and_debunks_2023}. They show that \add{prebunking} and debunking interventions \add{effectively change} a person's agreement with the main claims of an article, their credibility assessment of an article, and their intention to \add{agree or disagree} with the article. They also find that \add{debunking is} more effective than \add{prebunking} in certain situations.

One instance of a debunking intervention focused on trust ratings is provided by Pennycook and Rand, who argue that incorporating trust ratings of laypeople into social media platforms is effective~\cite{pennycook2019crowdsourcing}. Their investigation showed that laypeople across the political spectrum could distinguish between mainstream media outlets and hyperpartisan or fake news sources. 

A more user-centered approach focused on nudging~\cite{10.1145/3290605.3300733} was explored by Bhuiyan et al., who developed a browser extension to nudge users to assess news credibility on Twitter~\cite{bhuiyan2018feedreflect}. Their mixed-methods approach showed that the browser extension helps people access news credibility more accurately. There has also been work on collaborative fact-checking. Hassan et al., for instance, investigated an online fact-checking community on Reddit~\cite{10.1145/3308560.3316734}. A team of moderators and users in the community identify and verify check-worthy facts. Hassan et al.~describe how automated argument classification and stance detection tools can support such teams. For this, they systematize the process that image verification experts use to debunk or verify visual media.

Finally, the possibility of crowdsourcing the assessment has been explored. Venkatagiri et al., for instance, showed that the ratings from laypeople can be leveraged for this. They envisioned a system to facilitate the collaboration between journalists and \final{crowdworkers} in identifying manipulated images online~\cite{10.1145/3359209}. However, prior work showed that providing corrections \add{in crowdsourcing contexts} is complex. Corrections from friends were shown to be more effective than corrections from strangers~\cite{margolin2018political,hannak2014get}. Being corrected by human-looking bot accounts, however, is less effective and was shown to decrease the quality of retweets~\cite{mohsen_2021}. It leads to situations where less trustworthy sources are shared. \add{A central novelty of our research is to examine 1. what interventions people from different research fields propose in in-depth, qualitative interviews and 2. whether established misinformation experts support these interventions. The model we developed provides a basis from which to ideate potential new socio-technical interventions.}

\section{Methods}

\begin{table}[]
    \centering
    \small
    \setlength{\tabcolsep}{3pt}
    \color{black}
    \begin{tabular}{r|ccccccccc}
    \toprule
    & BIO-1   &   BIO-2   &   COM-1   &   COM-2   &   COM-3   &   CS-1    &   CS-2    &   CS-3    &   CS-4    \\
    \midrule
    Gender & f   &   m   &   m   &   m   &   f   &   m   &   f   &   m   &   f   \\
    Ph.D. & \checkmark  &   \checkmark  &   \checkmark  &   \checkmark  &   \checkmark  &   \checkmark  &   \checkmark  &   \checkmark  &   \checkmark  \\
    Prof. & \checkmark  &   \checkmark  &   \checkmark  &       &       &   \checkmark  &       &       &       \\
    \bottomrule
    \toprule
    & LAW-1   &   LAW-2   &   LAW-3   &   LAW-4   &   PHIL-1  &   POLSCI-1    &   POLSCI-2    &   POLSCI-3    &   POLSCI-4    \\
    \midrule
    Gender & f   &   m   &   m   &   m   &   m   &   m   &   m   &   m   &   f   \\
    Ph.D. & \checkmark  &   \checkmark  &   \checkmark  &   \checkmark  &   \checkmark  &   \checkmark  &   \checkmark  &   \checkmark  &   \checkmark  \\
    Prof. & \checkmark  &   \checkmark  &   \checkmark  &   \checkmark  &   \checkmark  &       &   \checkmark  &       &   \checkmark  \\
    \bottomrule
    \toprule
    & PSY-1   &   PSY-2   &   PSY-3   &   PSY-4   &   SOC-1   &   SOC-2   &   SOC-3   &       &       \\
    \midrule
    Gender & f   &   f   &   f   &   m   &   m   &   f   &   m   &       &       \\
    Ph.D. & \checkmark  &   \checkmark  &   \checkmark  &   \checkmark  &   \checkmark  &   \checkmark  &   \checkmark  &       &       \\
    Prof. &    &   \checkmark  &       &       &   \checkmark  &       &       &       &       \\
    \bottomrule
    \end{tabular}
    \caption{\add{An overview of domain experts, their field of expertise, their gender, and whether they were professors at the time of the investigation. For the domains, we use the following abbreviations: biology~(BIO), communication and media science~(COM), computer science~(CS), law~(LAW), philosophy~(PHIL), political science~(POLSCI), psychology~(PSY), and sociology~(SOC).}}
    \label{tab:expert_demographics}
\end{table}

\subsection{Semi-Structured Interviews}

The primary motivation of the \add{in-depth, semi-structured interviews} with an interdisciplinary group of people was to capture diverse perspectives across \final{several research fields, including many that do not traditionally contribute to research on misinformation}.

In the interviews, participants were asked to ``imagine a system that supports users in the detection and evaluation of misinformation and related phenomena.'' For this question, the interviewer emphasized to participants that they should think of the most helpful and impactful solution rather than the easiest or simplest one. The interviewer explicitly encouraged participants to disregard concerns regarding the technical or financial feasibility of their proposals. The follow-up questions of the semi-structured interviews explored how the solutions would help people, what functionalities a system implementing a solution would have, who would use the system and why, and what kind of explanations the system would provide. The interviewees were also asked what potential they see in the solution they described and what risks they see in implementing the solution. 

\add{To capture a diverse set of interventions that people with different scientific backgrounds consider helpful against misinformation, we recruited an interdisciplinary group of people from \add{various} fields. We only interviewed postdocs and professors to ensure that each interviewee had sufficient scientific expertise in their respective disciplines.}

\add{At a large public university}, we recruited 25~interviewees (10 female) from eight fields, including biology (1 male, 1 female), communication and media science (2 male, 1 female), computer science (2 male, 2 female), law (3 male, 1 female), philosophy (1 male), political science (3 male, 1 female), psychology (1 male, 3 female), and sociology (2 male, 1 female). \add{Table~\ref{tab:expert_demographics} provides an overview of the different experts, their gender, and whether they are professors. We had to omit age to avoid the identification of interviewees. The scientific fields that we interviewed were selected because they study either individuals and collectives (biology, philosophy, psychology, and sociology) or information systems and social networks~(computer science, communication and media science) or because they investigate or shape the larger political context (political science, law).} 

All interviewees had a Ph.D. or Ph.D.-equivalent in their field. Thirteen interviewees were professors. Twenty-two of the interviews (88\%) were conducted in German when both \final{the} interviewer and interviewee were native speakers. Three interviews were conducted in English when the interviewee was a native English speaker\final{,} while the interviewer spoke English as a second language. The interviews were scheduled for 60 minutes. \add{All} interviews were conducted using a videoconferencing tool. The audio of the interviews was recorded and transcribed by a professional transcription service. IRB-equivalent approval was sought and granted by the responsible authorities. \final{Although local regulations do not require a formal ethics review, an independent advisor reviewed the research protocol to ensure that the research adheres to ethical standards and legal regulations.} Informed consent (in line with the European GDPR) was obtained from all experts.

The interviews were conducted in December 2020 during the COVID-19 pandemic. At the time of the interviews, Donald Trump had just lost the U.S. presidential election but lied that he was the winner and claimed to be the victim of voter fraud. Shortly before the interviews, X (then named Twitter) and Facebook started to display warnings next to certain controversial tweets by then-president Trump and others.

\add{Each interview was conducted using the questionnaire in Section~\ref{sec:questionnaire}.} We analyzed the transcribed interviews using thematic analysis~\cite{Braun2006}. \add{The paper's first author performed an open coding where initial codes were derived directly from the data. The first two research questions were the starting point of the categories. The interview questions were mapped to the research questions (as explained in the Appendix). The first author carefully read the material multiple times and moved back and forth over the entire dataset. For each research question, he performed an inductive coding where he assigned text segments to broad categories like ``labeling misinformation''. These categories were then subdivided into sub-categories like ``labels for news sources'' or ``labels for news articles''. In weekly meetings with a supervisor, these categories were refined and discussed. The supervisor would review the results, checking for clarity, plausibility, and fit of each coding. Disagreements were discussed and resolved unanimously. When compiling the paper, the first author followed axial coding principles to find additional connections between the codes and merged and split them as needed~\cite{corbin2014basics}.} 

\subsection{Online Survey With Misinformation Experts}
\label{sec:sampling_strategy_survey}

\add{To understand how misinformation experts perceive the interventions proposed by non-experts, we conducted a follow-up survey with internationally recognized misinformation experts.} We operationalized the term misinformation expert as somebody who has served on the program committee of a conference or workshop focused on misinformation, disinformation, \final{or} fake news. We collected the names of experts from the websites of the \textit{Multidisciplinary International Symposium on Disinformation in Open Online Media (MISDOOM)}. At the time of the investigation, the symposium occurred thrice (2019 in Hamburg, Germany, 2020 in Leiden, The Netherlands, and 2021 in Oxford, The United Kingdom). At the MISDOOM conference, those who are part of the program committee are authors who have published at the conference themselves. We also added the members of the program committees of two recent workshops on mis/disinformation that were held as part of the ACM Conference on Human Factors in Computing Systems~(CHI): 1.~the \textit{Workshop on Opinions, Intentions, Freedom of Expression, ... , and Other Human Aspects of Misinformation Online} (CHI~2021) and 2.~the \textit{Workshop on Designing Credibility Tools To Combat Mis/Disinformation: A Human-Centered Approach}~(CHI~2022). The CHI workshop proposals and the program committee are curated, and the process is highly selective.

To recruit misinformation experts for the survey, we compiled a list of 102 experts who have published on misinformation and relevant phenomena. We excluded ourselves as authors of this study. We were able to retrieve the contact details of 92 misinformation experts via a Google Search for each expert's name and affiliation. We sent an online survey to these 92 experts. Thirty-four (34)~experts filled out the \add{complete} survey and stated that they answered all questions as described. Fifteen (15) participants were female (44\%), 16 male (47\%), and 3 (9\%) chose not to disclose their gender. The mean age of experts was 37.17 years (SD=5.91). We used the top-level domain of users' e-mail addresses as an indicator of what country they are residing in. Based on this information, the misinformation experts who responded reside in at least 14 countries on five continents. Continents include Australia, South America (Brazil), North America (Canada and the United States of America), Asia (Israel and South Korea), and Europe (England, Finland, Germany, Ireland, Italy, Poland, Romania, and the Netherlands). Every expert had at least a Master's degree. 85\% of experts had a Ph.D. The experts came from different fields. Eleven experts (5~male, 5~female, 1~undisclosed) had a background in computer science~(CS) or related fields like artificial intelligence, natural language processing, or human-computer interaction. Eight experts (5~male, 3~female) had communication and media science~(COM) expertise. Seven experts (4~male, 2~female, 1~undisclosed) had expertise in sociology, especially computational social science, and 2 (1~male, 1~female) had a political science and political communication background. One female lawyer also participated. Four experts (1~male, 2~female, 1~undisclosed) stated expertise in other fields, including cybersecurity, national security, linguistics, and media literacy.

\add{The survey was based on the answers provided by the domain experts in the interviews. For each potential intervention the domain experts suggested, the misinformation experts rated their agreement or disagreement. In the recruitment mail, we told misinformation experts that the goal of the survey was to validate the proposals from the interviews and that they would do this by rating the different solutions. The ratings were on a 5-point Likert scale with the answer options ''Strongly disagree'', ``Disagree'', ''Neither agree nor disagree'', ``Agree'', and ``Strongly agree''. The statements always followed the schema ``X can help people deal with disinformation.'' Examples include ``Education in schools and universities can help people deal with disinformation'' and ``Tools that find sources for claims can help people deal with disinformation.'' Misinformation experts were told that this is not a ranking task and that they can rate everything as ``Strongly disagree'' or everything as ``Strongly agree''.}

Considering the terminological confusion around misinformation and related phenomena, we \add{deliberately} asked survey respondents about disinformation to highlight the impact of intentionality. As discussed in the previous section and as shown by Altay et al.~\cite{altay2023survey}, the term misinformation frequently includes such intentionality. We, therefore, decided to use the term misinformation throughout the paper.

\section{Results}

In the following, we present our findings on the interventions against misinformation that an interdisciplinary group of people proposed in interviews~(RQ1), what potential risks they associate with these interventions~(RQ2), and how these proposals were perceived by misinformation experts~(RQ3).

Throughout the paper, we refer to the different interviewees by their field and an identifying number. We use the following abbreviations: biology (BIO), communication and media science~(COM), computer science~(CS), law~(LAW), philosophy~(PHIL), political science~(POLSCI), psychology~(PSY), and sociology~(SOC). CS-1, for instance, is computer scientist number one.

\subsection{Interventions Against Misinformation~(RQ1)}

In the following, we describe the interventions proposed in in-depth, semi-structured interviews. We thematically grouped the interventions: 5.1 Educating People, 5.2 Labeling Misinformation, 5.3 Providing Evidence, and 5.4 Deleting Misinformation.

\subsection{Educating People}

Interviewees like CS-1, CS-4, PHIL-1, and PSY-4 believe that awareness about misinformation should be increased. PHIL-1 thinks that this awareness is more important than any other approach. PSY-4 believes an awareness campaign is needed to show that misinformation is a problem. \add{CS-1 thinks such awareness would entail understanding that only some things posted on social media are true}. He thinks this could be accomplished through campaigns, e.g., \final{through} \add{government financing}.

\subsubsection{Promoting Media Literacy}

The importance of promoting media literacy, in general, was recognized by interviewees across domains~(COM-3, CS-2, CS-4, POLSCI-1, POLSCI-4, PSY-4, SOC-1, and SOC-2). CS-4 thinks people \add{should} be encouraged to proactively and independently fact-check claims for long-term impact. 

\add{These findings relate} to calls for critical thinking frequently mentioned by others~(LAW-1, LAW-2, PHIL-1, and SOC-2). LAW-1 and LAW-2 describe critical thinking as the ability to question information and think about whether something is plausible. 

However, the effectiveness of interventions focused on media literacy is not universally endorsed. LAW-3 criticizes the call for better training or more education as ``too generic''. He perceives this call for education as a ``pseudo-magic bullet''. He thinks focusing on the individual and teaching everyone how to handle misinformation is unrealistic. COM-1, who worked extensively on misinformation, reported that he is sometimes ``a bit worried that people say everything will work out if we just make people a bit more media-savvy.'' 

Interviewees also recognize the importance of promoting quality journalism and increasing trust in quality journalism \add{and helping people understand how journalism works}~(COM-1, COM-2, COM-3, LAW-2, LAW-3, POLSCI-2, and SOC-1). POLSCI-1, SOC-1, and COM-1 believe that a deeper understanding of how news is produced would help people to better assess news. \add{SOC-1 believes that established quality media needs transparent, open structures. These efforts include ``disclosing'' the whole process, from source to news report, including how information is checked and how journalists define what is newsworthy.  In this context, LAW-1, LAW-2, LAW-3, and COM-2, commented on the importance of trusted brands. More generally, COM-1 believes trust in important societal actors like the media is necessary.}

\subsubsection{Education in Schools and Universities}

A particularly salient context in which media literacy might be taught is within formal educational contexts, e.g., schools and universities.
Many interviewees commented on \add{education's role} in these settings~(COM-2, COM-3, CS-2, POLSCI-1, POLSCI-4, PSY-4, PSY-4, and SOC-2). COM-3 and CS-2 believe that training in media literacy should be compulsory in general education. POLSCI-1 thinks that training in media competency should be integrated into school life or at least taught at universities. Several interviewees~(CS-2, CS-4, POLSCI-4, PSY-3, and SOC-2) believe that especially children should be targeted, e.g., in dedicated school subjects or specialized courses\add{,} to inform them about fake news~(CS-4). \add{PSY-4 thinks education is important, especially computer science and social media education}. Other interviewees suggest that schools should encourage people to compare different sources~(PSY-3) and to critically examine a source and its agenda~(SOC-2). POLSCI-4 argues that the general level of education should be so high that people can contextualize the information and critically assess whether something is misinformation. PSY-4 said that it is important that schools teach the difference between opinion\final{s} and fact\final{s}. He proposes a certification that proves that somebody received special training in media competency. Based on her teaching experience at universities, COM-3 reports that even university students \add{cannot} evaluate which source is trustworthy and which is not. The call for education can also be found in the literature, e.g., regarding the cognitive resources it costs and the importance of presenting content in a worldview-affirming manner~\cite{doi:10.1177/1529100612451018} \add{and prompting social media users to think about the accuracy of the information they share or interact with~\cite{pennycook2021shifting}}. 

\subsubsection{Face-to-Face Conversations}

Interviewees also commented on the role of face-to-face conversations~(CS-1, CS-2, CS-4, LAW-2, and POLSCI-2). POLSCI-2, for instance, thinks that face-to-face conversations are the most important technique in the fight against misinformation. He argues that a ``personal'' conversation is always superior to ``impersonal'' technology. CS-4 thinks that people who believe in certain misinformation can only be convinced by their close social environment. POLSCI-2 recommends using a discursive language strategy with questions like: ``Where did you get this information from? Why do you believe [a piece of] information in this case that you might otherwise consider very unlikely?'' The goal should be to question the plausibility of the false information. During such conversations, one should not contradict people who believe in a conspiracy theory~(CS-2). \add{Instead}, one should try to understand their point of view and work from there. Instead of directly confronting the person who believes in misinformation or openly mocking them, CS-4 says it is important to show empathy and to convey to the person who believes in false information: ``\final{That's} ok, it can happen to me, too.'' 

\subsubsection{Crowdsourcing}

Another intervention proposed by interviewees is crowdsourcing the assessment. CS-4 envisions an online platform where users can ask their social network for help by putting certain misinformation and information up for discussion. The intervention is based on CS-4's insight that people may only be convinced by people close to them. CS-4 thinks such a system could create a discussion between people who know each other. 

\add{The crowdsourcing assessment proposal is most directly related to prior work on the crowdsourced assessment of fake news~\cite{sethi2017crowdsourcing} and trusted news~\cite{10.1145/3555637}.} 

\subsection{Labeling Misinformation}

In addition to interventions focused on Educating People, interviewees also proposed \add{several} socio-technical tools to label misinformation. The interviewees distinguished between labels for news sources and labels for content.

\subsubsection{Labels For News Sources}

\add{An} intervention that many interviewees commented on is labeling news sources~(COM-1, COM-2, COM-3, LAW-2, LAW-4, PSY-2, and PSY-3). LAW-2 and PSY-2 envision a database of websites that have been proven to misinform people. Other interviewees advocate highlighting reliable sources~(COM-3) and \final{high-quality sources}~(COM-2). Meanwhile, COM-3 warns that there is ``no objective map to classify different sources'' because this is always a ``political field'' and it is hard to classify sources in an unbiased way. COM-2 discussed the problems that arise when selecting trusted institutions. He recounts a fact-check run by Facebook where Breitbart was selected as a trustworthy source. Since he perceives Breitbart as highly controversial, COM-2 wondered how much sense it makes to select verified sources. LAW-4 believes that the ``verified badge'' on Facebook should be removed if somebody has posted ``too much fake news''. In contrast to those who believe that sources should be labeled, COM-1 believes that somebody with fundamental distrust cannot be convinced by adding a label. COM-2 is worried that such labeling can be viewed as too instructive or didactic. Rather than a warning, he believes \add{motivating} people to consume other content from verified sources \add{is better}. 
\add{Findings on the effectiveness of labels~\cite{10.1145/3415211,10.1145/3491102.3517717} and news reliability criteria~\cite{10.1145/3635147} could serve as a starting point to integrate the recommendations by the domain experts. The domain experts in our investigation do, however, add the important insight that there is no objective map to classify different sources.}

\subsubsection{Labels For News Articles}

In addition to labeling sources, interviewees recognize the potential of labeling content like individual news articles~(COM-1, CS-1, and PSY-2). PSY-2 believes labeling content like tweets as potentially suspicious could motivate people to investigate whether the information is actually true. CS-1 believes that such flagging of content could be implemented automatically by social media platforms. However, he thinks that the government would have to force the platforms to do so by law or regulation. POLSCI-3 thinks this might limit the spread of misinformation, although he is unsure how successful this would be. \add{However, he believes} this could encourage people to question a news story. SOC-3 is more skeptical. He doubts that labels on platforms like Twitter are helpful. He argues that ``if someone believes what [a particular leader known for lying in public statements is] saying, these labels will not change their mind''. For COM-1, labeling content and adding additional information is a good alternative to deleting or ``demetrifying'' content. 

The idea of labeling content relates to a large number of interventions focused on the automated detection of fake news~\cite{10.1145/3137597.3137600, 10.1145/3305260, perez-rosas-etal-2018-automatic,wang-2017-liar}. 
Kirchner and Reuter, for instance, \add{show} that users appreciate warnings, especially if they come with explanations~\cite{10.1145/3415211}. \add{These findings connect} to Epstein et al.~\cite{epstein2022explanations}, who showed that such labels increase sharing discernment. They also find that explanations increase the effectiveness of warnings. \add{This, again, could serve as a starting point to develop new ways of automatically flagging content and encouraging people to question a story.}

\subsection{Providing Evidence}

In addition to interventions focused on educating users and labeling misinformation, the interviewees also discussed several socio-technical solutions focused on finding evidence that helps users understand whether a source, a piece of information, or a claim is reliable or not.

A surprising finding is that unlike prior work on how to automatically detect misinformation~\cite{10.1145/3137597.3137600}, the interventions in this paper do not attempt to decide whether something is reliable, i.e., the interviewees proposed tools that can support people in their decision-making in ways that do not suffer from the limitations of the tools criticized by Asr et al.~\cite{doi:10.1177/2053951719843310}. The approaches described in this paper are more user-centered, i.e., they do not try to automate the task but help users deal with misinformation.

\subsubsection{Finding Sources For Claims}

\final{Interviewees also proposed} helping users find a source for \add{specific} claims~(LAW-1, LAW-3, POLSCI-3, and SOC-1). LAW-1, for instance, argues that deciding whether something is reliable should be based on finding sources for a claim. LAW-2 also thinks that if a source for a claim is provided, a user can assess whether this source is trustworthy based on what the user knows. BIO-1 imagines a system that verifies claims by referring to trustworthy sources. POLSCI-4 envisions a system \add{retrieving} scientific studies or evidence to prove or disprove a claim. She believes that such scientific evidence has high verisimilitude. LAW-3 proposes an automatic system that could decide whether something is reliable by finding sources for claims. In this context, POLSCI-3 cites Wikipedia as a model for \add{providing sources for claims}. The call to \add{offer sources} to claims relates to Haughey et al., who found that journalists who correct misinformation lack reliable data to substantiate decisions and decide what to report on and what not~\cite{10.1145/3415204}.

\subsubsection{Finding Information About Claims}

\add{Many} interventions are focused on finding information about claims~(BIO-1, COM-2, CS-2, CS-3, POLSCI-2, POLSCI-4, and PSY-2). COM-2, for example, argues that helping people search for articles or videos that contextualize an opinion or an event is important. Like COM-2, CS-3 describes a technical system that---using ``machine learning and very good searching''---could find a citation for ``everything you say''. CS-2 refers to this as the competency to ``read up on particular aspects''. This involves verifying a claim and the skill to look for a second source for a claim. She also refers to skills like skimming an article and identifying the central arguments and statistics worth checking. CS-2 envisions a technical system for finding relevant scientific studies or explanatory YouTube videos. To illustrate her point, she uses the false claim that face masks against COVID-19 do not allow people to breathe. She believes it would be best if a system could find a study showing that breathing with a mask is not a problem. LAW-2 and BIO-2 envisioned a chatbot-like system \add{allowing users to} ask questions in natural language. A user would enter a claim like ``Obama wasn't born in the United States\add{,}'' and the app would ``read up'' on the claims and provide a warning that says: ``Attention, this fact was disputed in another context.'' Regarding images, PSY-2 thinks it would be helpful to provide additional information about the \final{creator} of an image and the context in which it was taken. COM-2, for instance, mentioned reverse image search as a solution to understanding whether an image was taken out of context. In addition to that, CS-4 thinks that it would be helpful if a system \add{could} provide evidence about whether a photo is real or a ``deep fake''. The proposals in this context relate to work that showed that finding information about claims can be helpful~\cite{10.1145/3415204,doi:10.1177/1529100612451018}. 

\subsubsection{Finding Information About News Sources}

Interviewees also commented on the importance of finding information about a source~(COM-1, COM-3, POLSCI-2, and PSY-2). For POLSCI-2, the most important skill is \add{properly searching for information and assessing the quality of different sources}. COM-1 and POLSCI-2 think that people need to be able to understand where a piece of information comes from. PSY-2 believes that the intention and background of a source should be made transparent. This \final{would} enable people to determine whether someone is trying to deceive them. PSY-3 imagines a system that highlights the relative independence of ``political or other influences'' of a source. COM-3 imagines a quiz that teaches students how to gauge the quality of sources. In the quiz, users would rate whether a source is reliable and receive feedback as a learning experience. This could raise awareness of sources that may appear scientific but are not. Finding information about sources connects to Lewandowsky et al.'s recommendation to investigate whether a piece of information comes from a credible source~\cite{doi:10.1177/1529100612451018}. 

\subsubsection{Supporting Source Comparison}

Besides finding sources for claims and finding information about sources and claims, interviewees also advocated comparing different sources for the same claim~(COM-2, CS-1, LAW-2, and POLSCI-3). \final{LAW-2 argues that using multiple sources to verify a claim is important in journalistic practice}. CS-1 envisions a solution similar to online product comparison websites. He argues that if you want to buy a TV, you can use product comparison websites to compare the differences between TVs. He thinks that this comparison could be extended to information. COM-2 and POLSCI-3 highlight the importance of comparing different accounts and sources. COM-2 proposes a system that provides articles that are similar to a claim. However, considering the limitations of contemporary natural language processing systems \add{in computing} the similarity between articles, he is unsure how well this would work. He warns that such systems might only be able to find content that is similar in tonality. The solution connects to news aggregators like Google News, which can find and cluster different articles for a particular topic~\cite{das2007google,weaver2008finding,lee2015rise}.

\subsection{Deleting Misinformation}

The previous interventions focused on educating people, labeling misinformation, and providing evidence. A broad range of interviewees also advocated deleting misinformation~(COM-1, CS-2, LAW-3, LAW-4, POLSCI-3, POSCI-4, and SOC-3). Considering sources that mix \final{made-up information} with things that ``only have the goal to really harm our political system'', COM-1 does not understand why ``we are not discussing removing this from the Web''. Especially since COM-1 believes this type of misinformation is causing \add{much} damage. POLSCI-3 compares this to the moderation of comments. He thinks that certain kinds of content\add{,} like libel\add{,} need to be deleted. POLSCI-4 also thinks certain content should be deleted, e.g., if it is factually wrong or includes racist or discriminatory content. He refers to existing laws against hate speech in Germany, which allow victims of defamation to delete \add{specific} posts. He thinks that a duty to delete misinformation could be implemented in law. CS-2 also believes that one should think carefully about ``whether you really have to give people a platform when they spread conspiracies''. SOC-3 argues that governments could delete user accounts that publish misinformation, even though he believes this should only happen after warnings. LAW-3, however, believes that from a legal perspective, deleting content would require a comparatively complex evaluation process. He wonders which institution should decide whether something must be deleted or not. If Facebook, for example, would use automated tools to delete risky content, LAW-3 thinks it would be important to discuss the limits of Facebook's power. Considering proposals to delete misinformation~(COM-1, CS-2, POLSCI-3, and POLSCI-4), COM-2 warns that previous attempts at moderating or deleting content showed that those who publish extremist content migrate to other platforms. CS-2 thinks that if tweets or news stories are deleted, it would be important to receive details about why something was deleted. 

\subsection{\add{Interventions by Area or Field of Research~(RQ1)}}

\begin{table}[]
    \centering
    \small
    \color{black}
    \begin{tabular}{l|cccccccc}
        \toprule
        Interventions & BIO & COM & CS & LAW & PHIL & POLSCI & PSY & SOC \\
        \midrule
        Promoting Media Literacy               &   & \checkmark & \checkmark & \checkmark & \checkmark & \checkmark & \checkmark & \checkmark \\
        Education in Schools and Universities  &   & \checkmark & \checkmark &   &   & \checkmark & \checkmark & \checkmark \\
        Face-to-Face Conversations             &   &   & \checkmark & \checkmark &   & \checkmark &   & \\
        Crowdsourcing                          &   &   & \checkmark &   &   &   &   & \\
        \midrule
        Labels For News Sources                &   & \checkmark &   & \checkmark &   &   & \checkmark & \\
        Labels For News Articles               &   & \checkmark & \checkmark &   &   & \checkmark & \checkmark & \checkmark \\
        \midrule
        Finding Sources For Claims             & \checkmark &   &   & \checkmark &   & \checkmark &   & \checkmark \\
        Finding Information About Claims       & \checkmark & \checkmark & \checkmark & \checkmark &   & \checkmark & \checkmark & \\
        Finding Information About News Sources &   & \checkmark &   &   &   & \checkmark & \checkmark & \\
        Supporting Source Comparison           &   & \checkmark & \checkmark & \checkmark &   & \checkmark &   & \\
        \midrule
        Deleting Misinformation                &   & \checkmark & \checkmark & \checkmark &   & \checkmark &   & \checkmark \\
        \bottomrule
    \end{tabular}
    \caption{\add{This table illustrates the strong agreement on the proposed misinformation interventions across academic disciplines. It showcases how different fields converge on similar strategies and approaches to combat misinformation effectively.}}
    \label{tab:interventions_by_discipline}
\end{table}

\add{In the following, we will discuss the strong agreement on the proposed misinformation interventions across academic disciplines. Table~\ref{tab:interventions_by_discipline} arranges the findings for RQ1 by discipline. \final{Section~\ref{sec:example_responses} in the Appendix provides example responses and first author codes}. Most proposals came from the fields of computer science~(9 proposals) and political science~(9), followed by communication and media science~(8), psychology~(7), and law~(7). The table also documents that all fields have contributed to the set of interventions. Most intervention proposals are interdisciplinary. Experts from at least three fields independently proposed each intervention. There is one exception. The Crowdsourcing intervention was only mentioned by one computer scientist.}

\add{The Promoting Media Literacy intervention was proposed by the largest number of fields. Experts from seven fields supported the media literacy intervention. This support includes all fields except for biology. Experts from six fields supported the Finding Information About Claims intervention. Three more interventions were supported by experts from at least five different fields: Education in Schools and Universities, Labels For News Articles, and Deleting Misinformation. Table~\ref{tab:interventions_by_discipline} documents the broad agreement on the interventions across domains.}

\subsection{Potential Risks of Interventions~(RQ2)}
\label{sec:risks}

The interviewees \add{were also asked about the risks associated with the different intervention proposals}.

\subsubsection{Censorship}

\add{Various} interviewees discussed the consequences of their interventions for freedom of speech~(LAW-3, LAW-4, POLSCI-3). POLSCI-3 believes that any interference with information dissemination always poses the risk of becoming censorship, which is problematic since freedom of speech is a fundamental right in many \final{countries}~(LAW-3). LAW-4, therefore, argues that complex arrangements are needed to avoid ``collateral damage to freedom of opinion and information''. As examples of such complex arrangements, he described the different levels of regulation discussed in information law to avoid a conflict between freedom of opinion and ways of fighting misinformation, e.g., when deleting or labeling content. As an alternative to deleting content, COM-1 proposes downranking certain postings on social media platforms, e.g., by excluding them from recommendation or curation systems. The perceived advantage of downranking is that the impact of the misinformation is limited even though the information is still available. SOC-1 believes that governance solutions are needed to \add{uphold} the right to freedom of speech. 

A large body of work has addressed such governance questions~\cite{klonick2017new,10.1145/3313831.3376293,kleinsteuber2011control}. We will engage with these questions in-depth in the discussion.

\subsubsection{User Rejection}

The challenges associated with motivating people to use any of the interventions are recognized by \add{several} interviewees~(COM-1, COM2, CS-1, LAW-1, and POLSCI-1). COM-2, for instance, argues that earning user acceptance would be the most challenging aspect. POLSCI-1 believes that people do not have the time or \final{motivation} to check whether information is true. While COM-1 sees merit in media literacy, he believes training people to recognize misinformation would take a decade or longer. COM-2 is even more pessimistic. He argues that training users to verify sources is impossible. COM-2 also believes that users are not interested in explicitly and consciously reflecting on news sources. LAW-1 thinks that a lack of understanding could prevent people from using solutions, \final{e.g.}, because a tool might become too complex for the user. In addition, CS-4 warns that the proposed interventions could lead to an even larger bubble of distrust if a solution itself is viewed as ``fake''.

Interviewees also \add{discussed} the potential risk of only reaching those who already check information~(POLSCI-3) or question a situation~(PSY-2). PSY-3 argued those who do not want to question misinformation would not use the proposed interventions. \final{POLSCI-2 believes that only readers who are already critical of conspiracy theories are being reached}. SOC-1 worries that \add{his proposed interventions} are only reaching a more highly educated middle class\final{,} while ``many others remain excluded''. CS-4 argues that a basic awareness of the problem of misinformation is necessary to use solutions. PSY-4 thinks those prone to believe in misinformation are unlikely to use tools or look for additional information because they will have reservations. This risk connects to the problem that checking a piece of information against other knowledge is effortful and requires motivation and cognitive resources~\cite{doi:10.1177/1529100612451018}.

\subsubsection{Mistakes}

Another risk that interviewees like COM-2, CS-1, and CS-4 commented on are mistakes, e.g., due to the limited capabilities of a solution. COM-2 warns that even small mistakes by a system could lead to the system being perceived as unreliable. LAW-2 warns that especially algorithm-based approaches would have a risk of error. \final{Therefore, it would not be possible to automatically determine with certainty whether something is false}. LAW-2 connects this to the risk of overreliance. He thinks that the better the system, the more dangerous its mistakes. \add{An error can have a potentially large effect if a system is very reliable}. CS-2 also believes that the biggest problem is a ``false sense of security'' that people who use such a system could develop. In addition, BIO-2 thinks that misinformation disseminated by a fact-checking system would be more likely to be believed. COM-2 does not think it is possible to completely automate the process. While he recognizes great progress regarding the automatic moderation of hate speech, he thinks contemporary technology is not sophisticated enough to handle longer articles and comments. COM-2 considered the risk posed by possible indictments by people who believe to be wrongly accused of spreading misinformation. LAW-2 also mentioned the potential risk of emerging real news getting filtered at an early stage.

\subsubsection{Potential for Manipulation}

Another critical risk associated with the interventions is the potential for manipulation~(SOC-2) and the risk of manipulating people~(POLSCI-4). Regarding the Labeling Misinformation interventions, \final{CS-3 and POLSCI-4 are both concerned about what might happen if a malicious actor takes control of a solution}. SOC-3 worries that somebody will try to control solutions to gain power. He \add{fears} that labeling interventions will become ``very political\final{,} very fast''. LAW-3 believes that, in the wrong hands, solutions would be ``fatal''. He also mentions the risk of governments controlling information. LAW-3 discusses the power \add{of} those who develop a solution and the risks that this poses. He \final{highlights} how challenging it is to remain objective and neutral. He was also worried that people might question whether the developers of a solution are actually neutral.

More generally, PHIL-1 is afraid of what happens if misinformation \final{authors} or spreaders use solutions to disseminate misinformation, i.e., whether some or any of these potential interventions consistently bend users towards truth, like think-pair-share or peer instruction exercises have been shown to do in educational contexts~\cite{tullis2020does}, regardless of who is deploying them. In addition to the risk of being manipulated, CS-1 recognizes the risk of giving people the impression that they are being manipulated.

\subsubsection{Backfire Effect}

POLSCI-1, among others, believes it is problematic to correct misinformation in hindsight. He thinks this frequently does not work and refers to the backfire effect, where repeating a message to debunk it strengthens the original misinformation. POLSCI-1 believes correcting misinformation can \add{make people} look for information that fits their beliefs and perspectives more strongly. POLSCI-4 also assumes that hearing the same claim repeated by different media could lead people to believe it. POLSCI-2, therefore, thinks that the repetition effect is extremely important. These risks are recognized in the literature on misinformation, e.g., Greenhill and Oppenheim~\cite{greenhill2017rumor}. Lewandowsky et al. cite a large body of research that shows that repetition of information can strengthen that information in memory and ``thus strengthens belief in it''~\cite{doi:10.1177/1529100612451018}. However, Wood and Porter argue that evidence of a factual backfire effect is far more tenuous than prior research suggests~\cite{wood2019elusive}. \add{Their investigation does not find corrections capable of triggering a backfire effect}.

\subsection{\add{Potential Risks by Area or Field of Research (RQ2)}}

\begin{table}[]
    \centering
    \small
    \color{black}
    \begin{tabular}{l|cccccccc}
        \toprule
        Potential Risks & BIO & COM & CS & LAW & PHIL & POLSCI & PSY & SOC \\
        \midrule
        Censorship                             &   & \checkmark &   & \checkmark &   & \checkmark &   & \checkmark \\
        User Rejection                         &   & \checkmark & \checkmark & \checkmark &   & \checkmark & \checkmark & \checkmark \\
        Mistakes                               & \checkmark & \checkmark & \checkmark & \checkmark &   &   &   &   \\
        Potential for Manipulation             &   &   & \checkmark & \checkmark & \checkmark & \checkmark &   & \checkmark \\
        Backfire Effect                        &   &   &   &   &   & \checkmark &   &   \\
        \bottomrule
    \end{tabular}
    \caption{\add{This table details the risks associated with misinformation interventions as identified by the different disciplines. It highlights the strong agreement across disciplines regarding these risks, demonstrating a unified perspective on the potential challenges and considerations in combating misinformation.}}
    \label{tab:risks_by_discipline}
\end{table}

\add{Table~\ref{tab:risks_by_discipline} provides an overview of which risks were mentioned by \final{participants from} which field. Political science and law are the fields that refer to the most potential risks. Both referred to four of the five risks that we identified. Five of the eight fields mentioned at least three risks. Our interdisciplinary investigation also showed that the risk\final{s} of the interventions \final{are} discussed across disciplines. Experts from at least four fields discussed each risk. The only exception is the backfire effect, which was only brought up by political scientists. The risk most widely commented on was user rejection (discussed by 6 disciplines), followed by the potential for manipulation~(5), censorship~(4), and mistakes~(4). We again document an interdisciplinary consensus on the importance of the risks among experts from different fields.}

\subsection{Expert Perspective on Intervention Proposals (RQ3)}

\begin{figure*}
    \centering
    \includegraphics[width=.9\linewidth]{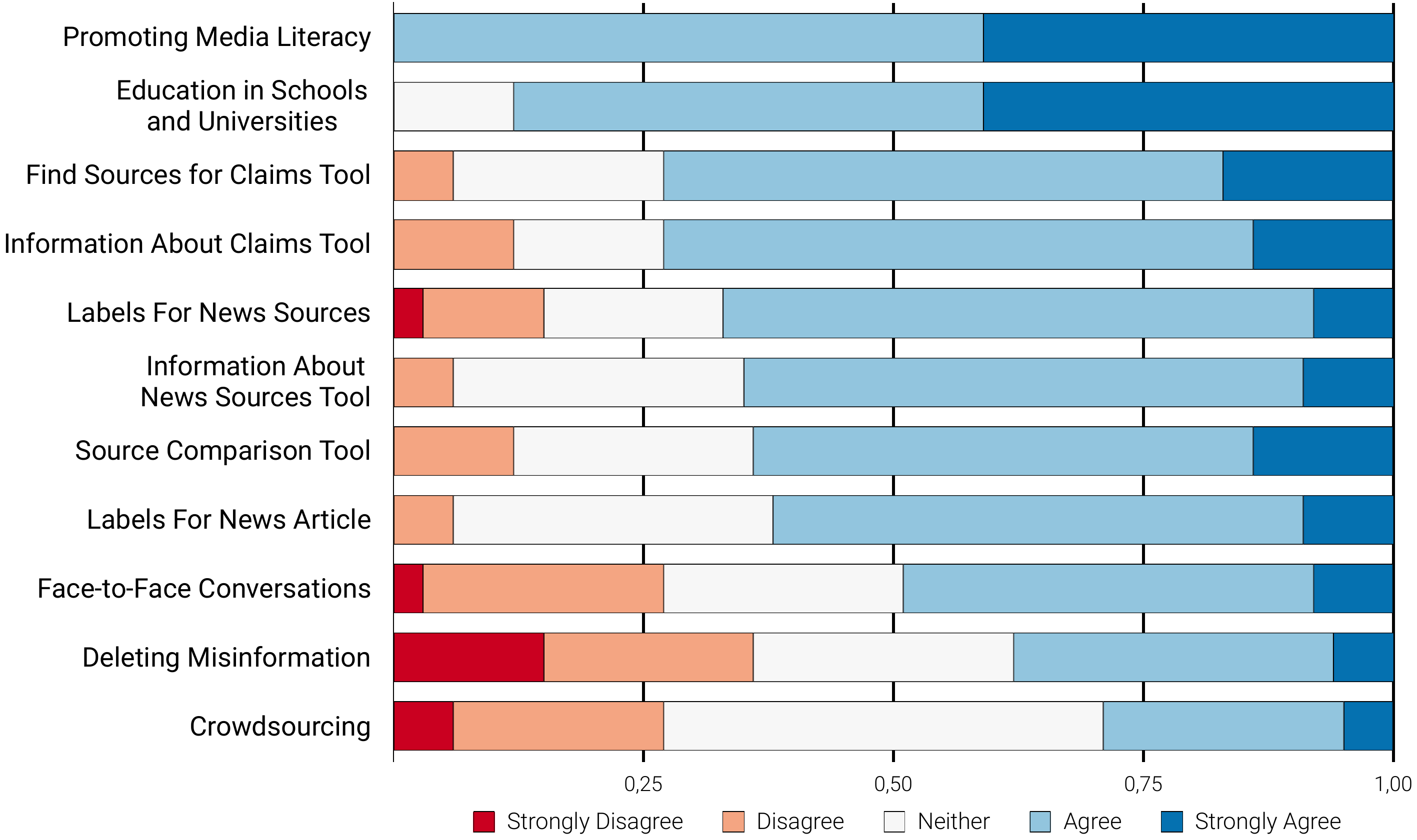}
    \caption{We conducted an online survey with 34~misinformation experts to understand how the intervention proposals from the interviews are perceived. The figure shows their agreement and disagreement to the questions like ``Promoting media literacy can help people deal with disinformation.''}%
    \label{fig:experts_approaches}%
    \Description{Bar charts of the agreement of misinformation experts to the intervention proposals made by interviewees. We see very high agreement for proposals like Promoting Media Literacy, Education in Schools and Universities, the Find Sources for Claims Tool, and Information About Claims Tool. All are supported by nearly three out of four misinformation experts and more. Face-to-Face Conversations, Deleting Disinformation, and Crowdsourcing are supported by less than half of misinformation experts.}
\end{figure*}

To include the perspective of current research, we surveyed 34~misinformation experts. We evaluated whether the misinformation experts agree that the interventions proposed by interviewees can help people deal with misinformation(Figure~\ref{fig:experts_approaches}).

All misinformation experts agree that promoting media literacy can help people deal with misinformation. Nearly nine of 10 misinformation experts (88\%) also believe education in schools and universities is helpful (12\% neutral). For these two interventions, it is also important to highlight the low disagreement. None of the misinformation experts disagreed that education in schools and universities and promoting media literacy is helpful.

Technical interventions proposed by the interviewees were also recognized as helpful. About three out of four misinformation experts agree that tools that find sources for claims (74\% agree, 6\% disagree) and tools that find information about claims (74\% agree, 12\% \final{disagree}) could, if successfully implemented, help people deal with misinformation. Labels that classify news sources as reliable or unreliable (68\% agree, 15\% \final{disagree}) were also perceived as helpful if successfully implemented. Almost two out of three misinformation experts also perceived tools that would find information about sources (65\% agree, 6\% disagree) or tools that enable comparing different news sources (65\% agree, 12\% disagree) as helpful. Still, 62\% of misinformation experts agreed that labels for news articles would be helpful (6\% disagree). Half of the interviewees (50\%) affirm that face-to-face conversations could help people deal with misinformation (26\% disagree). 

Only two of the interventions proposed by the interviewees were not perceived as helpful by \add{most} misinformation experts. The most controversial intervention was deleting misinformation. While 38\% of misinformation experts agreed that this is helpful, 35\% disagreed. The least helpful approach, as perceived by these experts, was crowdsourcing the assessment. Less than every third misinformation expert (29\%) agreed that this could help, while every fourth expert disagreed that this was a helpful intervention (26\%).

Nine of the eleven interventions proposed by various interviewees were perceived as helpful by a majority of the misinformation experts. Four interventions were supported \final{by two out of three experts}. This agreement indicates that the interventions presented here may be worth pursuing in future research.

\section{Discussion}

With this paper, we collect potential interventions in interviews~(RQ1) and discuss the risks associated with these proposed interventions~(RQ2). We also show how misinformation experts perceive the interventions proposed by non-experts \final{from different domains}~(RQ3). We found that interventions from the categories Educating People, Providing Evidence, and Labeling Misinformation were supported \add{by an interdisciplinary group of researchers from different domains.} We also identified Deleting Misinformation as controversial. Based on our insights, this discussion proposes the Phase Model of Misinformation Interventions and situates each intervention in the model. \add{This model can help practitioners decide what solutions against misinformation they should implement. The model is unique since many interventions against misinformation are designed and developed without considering this interdisciplinary perspective. We provide this interdisciplinary perspective and show which interventions are approved by misinformation experts, thus providing a starting point for interventions with broad appeal and support.}

\subsection{The Phase Model of Misinformation Interventions}

\begin{figure}
  \centering
  \includegraphics[width=\linewidth]{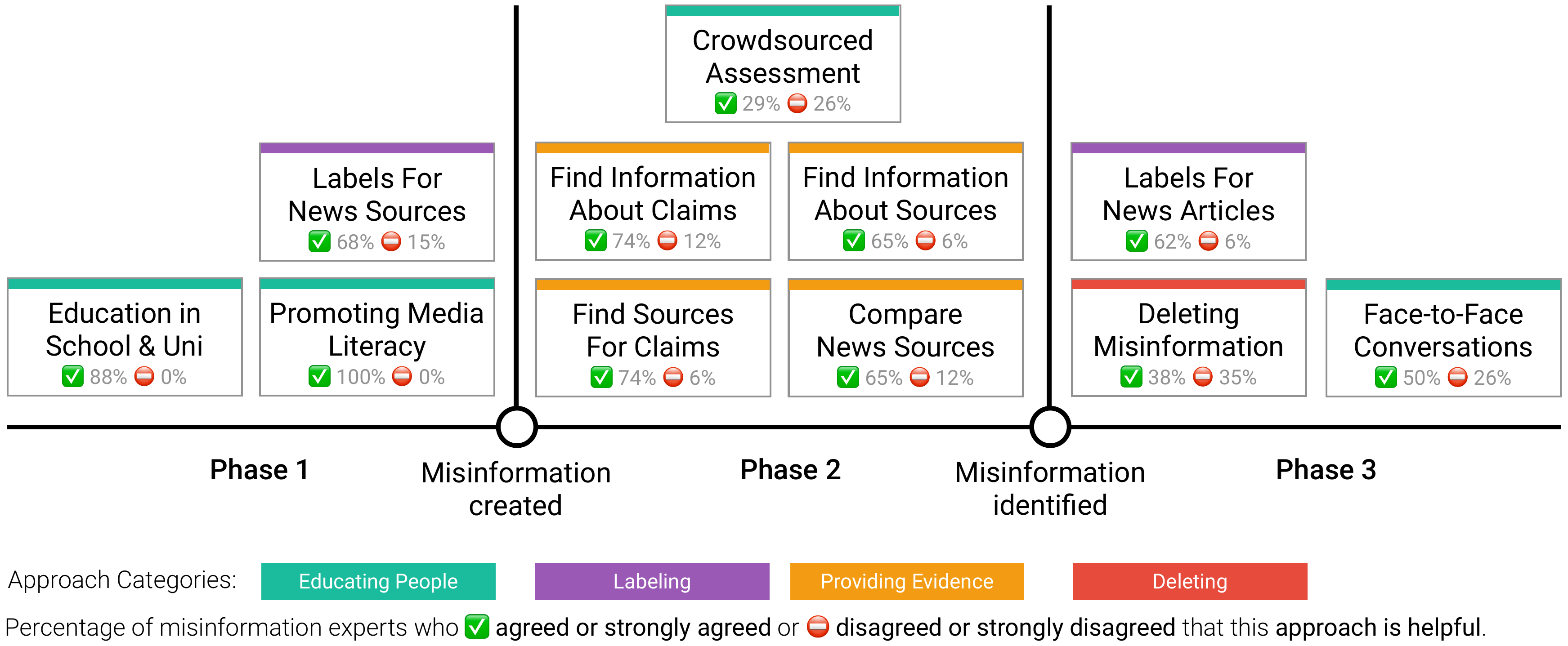}
  \caption{We distinguish three phases that interventions can target: before misinformation is created (Phase~1), before misinformation is identified (Phase~2), and after misinformation is identified (Phase~3). We situate the different interventions proposed by our interviewees and color-code the categories: Educating People (green), Labeling Misinformation (purple), Providing Evidence (orange), and Deleting Misinformation (red). We indicate how many misinformation experts agreed or disagreed that this intervention is helpful.}~\label{fig:approaches}
\end{figure} 

\add{Our paper is the first to empirically establish the three phases of misinformation interventions and how they can be best addressed by interventions.} Figure~\ref{fig:approaches} illustrates the \textit{Phase Model of Misinformation Interventions} \add{and the} three phases: before misinformation is created (Phase~1), before misinformation is identified (Phase~2), and after misinformation is identified (Phase~3). The distinction between Phase~2 and Phase~3 is necessary because a lot of misinformation remains undetected~\cite{brennen2020types} and because the identification of misinformation lags behind the creation of misinformation, typically by 10 to 20 hours~\cite{10.1145/2872518.2890098}. 

To create the Phase Model of Misinformation Interventions, \final{the paper’s first author carefully assigned each intervention to a phase by reading the interviewees’ descriptions and selecting the phase they deemed the best fit}. The \add{supervisor} then validated this assignment. \add{For the phase model, we considered the use cases where the intervention starts to become effective.} Promoting Media Literacy, for instance, is helpful for any of the three phases. However, in practice, we believe it is most effective for the setting right before misinformation is created. We, therefore, assigned it to Phase~1.

Based on the interventions proposed by interviewees and the ratings these potential interventions received from misinformation experts, we identify three Phase~1 interventions whose applicability is independent of a particular piece of misinformation. Such interventions can be deployed before misinformation is even created. The five Phase~2 interventions proposed by the interviewees are applicable after the misinformation was created and before it is identified. The \final{interventions} can help individuals to identify misinformation themselves. Finally, we also distinguish three Phase~3 interventions that are particularly helpful after misinformation has been identified as misinformation.

\subsubsection{Phase~1}
This phase includes non-technical interventions from the category \textbf{Educating Users}---\textit{Media Literacy} and \textit{Education in Schools and Universities}---and the socio-technical \textit{Labels For News Sources} approach. The latter only works if a misinformation source is known beforehand or if reliable sources are labeled. \final{Empirical research showed the effectiveness of digital media literacy interventions in practice. This includes research that shows the importance of information literacy in general~\cite{jones2021does} and for adult populations in particular~\cite{lee2018fake}, as well as empirical evaluation of Facebook tips that increase discernment between mainstream and false news~\cite{guess2020digital}.}

\subsubsection{Phase~2}
This phase entails interventions for the time between the creation of a misinformation story and the point in time where a consensus is reached that a story is misinformation. For Phase~2, the interviewees distinguish four interventions from the category \textbf{Providing Evidence}: the \textit{Find Information About Sources} tool, the \textit{Compare News Sources} tool, the \textit{Find Information About Claims} tool, and the \textit{Find Sources For Claims} tool. The phase also includes the socio-technical \textit{Crowdsourcing} intervention focused on enabling users to ask others. 

\subsubsection{Phase~3}
Here, we include interventions for the time after misinformation is identified. Specifically, the \textit{Labels For News Articles} as well as the \textit{Deleting Misinformation} interventions. In addition to that, the \textit{Face-to-Face Conversations} intervention from the category \textbf{Educating People} is included. \add{Solutions for Phase 3 are, again, focused on long-term \add{effects}. Labels and the deletion of misinformation are particularly sustainable as these actions have long-term \add{impacts}.}

\add{Our third research question was to understand how misinformation experts perceive the intervention from RQ1. The} survey showed that the \textbf{Educating People} interventions for Phase~1 are perceived as the most helpful by \add{misinformation} experts. While non-technical interventions can be found in all phases, they are perceived as less helpful for Phases 2 and 3. The \textit{Crowdsourcing} intervention (Phase~2) was judged \add{the} least helpful of all eleven interventions. The category \textbf{Providing Evidence} includes Phase~2 interventions that are perceived as helpful by the majority of misinformation experts. As the authors, we consider this category particularly noteworthy because these interventions are always available and empower users to identify misinformation. This is an important advantage over Phase~3 interventions like \textit{Labels For News Sources} and \textit{Deleting Misinformation} because it takes time to reach a consensus that something is considered misinformation. Our findings extend on prior work on the automatic labeling of misinformation~\cite{perez-rosas-etal-2018-automatic,wang-2017-liar,10.1145/3137597.3137600,10.1145/3377478,10.1145/3305260,10.1145/3274351} by \final{highlighting the additional information users need beyond individual labels}.

Overall, the discussion documents how the interventions proposed in the interviews support at least three distinct phases. We also find that misinformation experts perceive the proposals as helpful, especially Phase~1 and Phase~2 interventions like \textit{Media Literacy}, \textit{Education in Schools and \final{Universities,}} and \textit{Labels For News Sources}, as well as the four different tools in the \textbf{Providing Evidence} (Phase~2) category. 

\subsection{\add{Real-World Applicability of Phase Model for Misinformation Interventions}}

\add{A central idea of the phase model for Misinformation Intervention is to help practitioners decide which interventions to develop. The phase model and the interventions presented in this paper help them understand what \final{phase} misinformation intervention can target and how widely accepted the interventions are by different domains and misinformation experts. The phase model itself can also be taught to promote media literacy.}

\add{The phase model can help a variety of stakeholders. Stakeholders concerned about the potential impact of misinformation on elections might focus on Phase 1 solutions, which require a big investment but which could have a long-term impact by helping people recognize misinformation before they encounter it.}

\add{Other stakeholders may benefit from a deep understanding of Phase 2. Solutions for this phase are particularly relevant for stakeholders that expect to be subject to misinformation campaigns, e.g., a presidential candidate for a specific party or a CEO. Considering the solutions identified in our investigation, these stakeholders could ensure that they monitor what information is published about them and by whom. They could then strategically prepare material that helps users find information about claims and sources for claims.}

\add{In Phase 3, stakeholders may be worried about a particular piece of misinformation like Pizzagate. Stakeholders in a situation where the misinformation is already created might benefit more from helping users Find Information About Claims or Finding Sources For Claims.}

\subsection{Interventions \& Anticipated Risks}

In the following, we will discuss which of the interventions (Figure~\ref{fig:approaches}) are connected to which of the anticipated risks (Section~\ref{sec:risks}), informed by our understanding of the interviews. Overall, we believe none of the interventions are free of risk. At the same time, some risks\add{,} like censorship\add{,} are more dangerous than others, like the risk of users rejecting any particular intervention. We find that all except for one potential intervention may avoid the risk of censorship; that exception is the Deleting Misinformation proposal. \add{This risk could explain why the misinformation experts we surveyed ranked the Deleting Misinformation intervention comparatively low}. We believe that this low agreement also coincides with how prone deletion-based interventions might be to mistakes and manipulation.

We believe that interventions focused \final{on increasing} users’ knowledge about misinformation, providing labels, or deleting misinformation may not suffer from user rejection as much as others because other interventions may require additional user engagement and mental energy that the user would need to expend or tools that the user would need to learn how to wield. \add{However, prior work did report on users' negative attitudes towards removing} misinformation~\cite{10.1145/3411763.3451807} and adding labels~\cite{10.1145/3555637}. Labels for individual news articles are connected to the backfire effect since the strong wording or the visuals \add{could} potentially reinforce users' belief in misinformation~\cite{lewandowsky2012misinformation}.

The two potential interventions that we anticipate having the lowest risks are also the most highly ranked: Media Literacy and Formal Education. We believe their risk is comparatively limited because they prepare people to encounter misinformation and are not specific to individual misinformation stories. However, literacy training could be manipulated to push a political agenda. Regarding mistakes in identifying misinformation and manipulation to get around these potential solutions, such education-based interventions may still be less affected by mistakes made by those who provide the solutions or efforts to manipulate the results. The user\final{s} of such interventions would be taught generic strategies rather than passively receiving decisions made on their behalf.

\subsection{Epistemic Assumptions}

Our investigation revealed a strong potential for technical interventions for Phase~2, i.e., after misinformation is created and before misinformation is identified. We believe that interventions like building tools for \textit{Finding Information About Claims} and \textit{Finding Sources About Claims} are worth exploring. At the same time, further work is needed to explore how valid the epistemic assumptions behind these interventions are. The implicit assumption that all misinformation can be integrated into a dialectical framework with a thesis (the misinformation) and an antithesis (the correction) may be misguided. Consider an article like ``BREAKING: Nancy Pelosi's Son Was Exec At Gas Company That Did Business In Ukraine'', which was one of the most popular misinformation stories in 2019 with more than 8.5 million views~\cite{avaaz_us_2019}. It may be difficult or impossible to find sufficiently convincing evidence against such a claim because it is difficult to prove a negative. There may be little incentive to publish stories that claim the opposite, i.e., that Nancy Pelosi's son has never worked in Ukraine. The same epistemic limitations apply to the story CS-2 mentioned regarding people who allegedly cannot breathe while wearing masks against COVID-19. CS-2 proposed a solution that automatically finds a scientific study that shows that breathing with a mask is not a problem. The question is: Why would such a study get published? Another related question is: Who would fund such research? For many made-up stories, especially stories where claims are framed as allegations or allusions, there may not even be an antithesis to a misinformation thesis. Considering that some misinformation \final{claims are} \add{more challenging} to refute than others, we believe that interventions in the \textbf{Providing Evidence} category may only be a part of the solution.

\subsection{\add{Contextual \& Platform-Specific Interventions}}

\add{Our investigation showed that interventions against misinformation are contextual. This insight motivated us to consider that certain interventions mentioned in this paper must be applied differently based on where and how users encounter misinformation. What interventions are applicable depends on whether misinformation appears in a post, if the post is quoting a news article, or if it shows up when people search for information on a search engine, etc. We, therefore, encourage practitioners to think strategically about what intervention to add where. For instance, if an individual user shares misinformation about the origin of the COVID-19 virus on a social media platform, you can either label all posts by the user (akin to Labels for News Sources) or the individual post on the social media website (akin to Labels for News Articles). In this case, it would not be appropriate to label the entire social media platform or to encourage people to \final{question} the \final{social media platform} as a \final{whole}. In addition to being contextual, the interventions must be adapted to the specific platform context. In the context of the social media site \final{Reddit}, it might, for instance, also be necessary to focus on individual subreddits as sources.}

\add{Another platform-specific aspect relates to how information is presented to users. Social media platforms primarily used on mobile devices may offer limited space for displaying detailed information since the screens are comparatively small. In this case, it might be preferable to provide short information like ``This source has published 42 known cases of misinformation'' in the interface and show additional information only where necessary or where the user actively selects it.}

\subsection{Design and Research Recommendations}

\add{Our unique contribution is a set of interventions with broad approval across disciplines. Misinformation experts strongly support these interventions. We can make concrete design and research recommendations for future work based on these insights. Considering the distinct phases we identified, we show why the fight against misinformation requires more than one unified solution.} 
\add{The diversity of findings we identified indicates that combining different interventions may be crucial to support users in the fight against misinformation. Our results add important interdisciplinary insights to existing proposals\final{. We} are also the first to provide an actionable overview of which intervention to pursue. Based on the empirical insights presented in this paper, we believe that when fighting misinformation, the whole is greater than the sum of its parts}. \add{Interventions against misinformation are like the Swiss cheese model~\cite{reason2000human}: each intervention has weaknesses, but when layered together, they dramatically cut down on the failure rate.} 
For maximal protection against misinformation, Phase~1 interventions that can be applied before misinformation is created should be combined with interventions from Phase~2 and Phase~3. For example, education about how journalism works (Phase~1) would help people understand and use tools that find and understand information about claims and sources (Phase~2).  
In particular, the three interventions discussed in the previous section---\textit{Labels For News Sources} (Phase~1), \textit{Finding Information About Sources} (Phase~2), and \textit{Deleting Misinformation} (Phase~3)---could all be integrated into social media platforms like Facebook, YouTube, X, BlueSky, Mastodon, TikTok, Telegram, WhatsApp, and others. 
\add{We think implementing more than one intervention can lead to a virtuous circle where implementing one intervention strengthens other interventions. The recommendation to label news sources, combined with tools that help users find information about a source and better media literacy training in formal education and beyond, is a potent combination}.

The combination of Labels For News Sources (Phase~1), Finding Information About Sources (Phase~2), and Comparing News Sources (Phase~2) interventions have several advantages over other combinations. First, they are scalable interventions. Providing Labels For News Articles could lead to a Don Quixotesque fight against windmills; the cost of registering a domain and building an audience to subvert Labels For News Sources is significantly higher than spreading individual misinformation stories. Second, conditioned on (1) the existence of a trustworthy \add{and independent} news-labeling institution and (2) the user selecting such an institution over any untrustworthy alternatives, a clear news source label from a trustworthy institution is actionable and may reduce the cognitive load of the news-consuming user. \add{This, however, is contingent on the existence of such an institution. Potential examples \add{of} such institutions are public broadcasting services like the BBC in the United Kingdom and the ARD in Germany, who employ well-trained and independent journalists. Wikipedia also already classifies news sources' reliability and might be considered a provider of intersubjective agreement on the reliability of sources. Finally, NewsGuard already provides such labels as a service. Our findings provide a starting point for a contextual inquiry into the user acceptance of \final{such} labels.}

From a socio-technical perspective, these Phase~1 and Phase~2 interventions may have the most impact. Therefore, we strongly encourage platform providers to develop and evaluate the effectiveness of such tools in their platforms as soon as possible. 
\add{In parallel, civic hackers and journalistic organizations could fill the gap by creating tools, websites, and browser extensions that \final{implement interventions}.}

Regarding Phase~3 interventions, \add{we} believe that some or all indisputable cases of misinformation should be deleted or deplatformed. \add{This is informed by the} illegality of Holocaust denial, i.e., the act of denying the Nazi genocide of Jews in the Shoah, in sixteen European countries (including Germany) and Israel~\cite{enwiki:1011129881}. Laws against this and other indisputable facts could be used as a legal basis to delete misinformation. The legal perspective already recognizes crimes and offenses that directly apply to misinformation. Klein and Wueller, for instance, cite defamation, intentional infliction of emotional distress, unfair and deceptive trade practices, criminal libel, and cyberbullying as relevant crimes and offenses~\cite{klein2017fake}. Deleting obvious lies or libel could be used to fight a certain, obvious subset of misinformation with an established, transparent process that does not put freedom of speech at risk. Legal aspects related to misinformation are also discussed in recent laws like the Digital Services Act~\cite{savin2021eu} or the EU Code of Practice on Disinformation~\cite{kuczerawy2019fighting}. 

\subsection{Limitations \& Future Work}

The most important limitation of our approach lies in the cultural and socio-economic backgrounds of the interviewer and the interviewees. We are aware of their impact and tried to highlight them throughout the paper by referring to participants' statements based on their expertise, e.g., CS for computer science or COM for communication and media science. Due to our focus on Western democracies, the applicability of our findings and potential interventions in other settings, such as totalitarian regimes or countries with vastly different media landscapes, requires further research. We \final{also} recognize that various social and cultural factors influence preferences for or against specific interventions. The reasons and interventions proposed by the German interviewees are likely influenced by their positionality and socialization. \add{This influence} includes, among other things, the legal situation in Germany and the European Union and the historic role that propaganda has played in Nazi Germany and the German ``Democratic'' Republic. Considering the influence of cultural contexts, our methodology could be used to do similar investigations in other countries.

\section{Conclusion}

\add{Our paper is the first to provide an actionable, interdisciplinary overview of} interventions against misinformation~(RQ1). We categorized these interventions into four categories: Educating People, Providing Evidence, Labeling Misinformation, and Deleting Misinformation. We discuss the potential risks associated with these interventions~(RQ2) and show that an international team of misinformation experts supports the helpfulness of a majority of interventions~(RQ3). In the discussion, we organize the interventions into the \textit{Phase Model of Misinformation Interventions}. We hope the insights presented in this paper can empower researchers and practitioners to make the world more impervious to misinformation.

\begin{acks}
I would like to thank Elena L. Glassman for her valuable contributions to the early conception and development of this work. Her insights and guidance were instrumental in shaping this manuscript. I thank all participants for their time and their insights. I also thank the reviewers for their helpful and constructive feedback.
\end{acks}

\bibliographystyle{ACM-Reference-Format}
\bibliography{references-interviews}

\appendix

\section{\add{Questionnaire}}
\label{sec:questionnaire}

\add{In the in-depth, semi-structured interviews, we asked the domain experts the following questions:}

\begin{enumerate}
\item How should one support people to recognize disinformation?
\item Now please imagine a system that supports users in the detection and evaluation of disinformation. How would the system help people? What functionalities does the system have?
\item Who would use the system and why? 
\item In \final{what} situations and contexts would such a system be used?
\item What explanations would the system offer? 
\item Why do you think these explanations would be helpful?
\item What potentials do you see in using such a system?
\item What risks do you see in using such a system?
\end{enumerate}

\add{Interview questions 1-7 were used to elicit an in-depth perspective on RQ1 on interventions against misinformation. Question 8 was a follow-up question focused on the potential risks of deploying the interventions (RQ2).}

\section{\final{Example Responses and First Author Code}}
\label{sec:example_responses}

\subsection{\final{Promoting Media Literacy}}

\begin{itemize}

\item \final{POLSCI-1: ``Building higher media literacy in society. The idea is that people are able to filter information themselves. This would result in them having more trust in their own competence, but also in the media overall.'' (Media Literacy and Media Education)}

\item \final{COM-1: ``So, there is the simpler part, which can actually be solved by better explaining to people, `Where does this information come from? How do journalists work?' It is the whole portfolio of solutions to explain media, information, and journalism, that can be applied. Some of these were already used pre-Internet, but now must also be applied to platforms and intermediaries, so that people say, `Aha, then I am actually more skeptical'.'' (Media Literacy and Media Education)}

\item \final{BIO-1: ``And then actually making it clear to people what basic principles [of journalism] are that they can rely on. But defining those is, I believe, not easy.'' (Media Literacy and Media Education)}

\end{itemize}

\subsection{\final{Education in Schools and Universities}}

\begin{itemize}

\item \final{PSY-3: ``Encouraging [people in] school education and similar settings to question sources more critically, compare them, and, yes, perhaps trust certain sources more than others.'' (Media Literacy and Media Education)}

\item \final{POLSCI-1: ``In principle, it would be more of a didactic approach, where you would somehow --- I don’t know --- integrate it into everyday school life. To gradually build up media literacy. Or in university studies. Or somewhere else.'' (Media Literacy and Media Education)}

\item \final{POLSCI-4: ``For example, raising awareness in the classroom for children could help --- I speak from experience here. My children, when they were in school in Boston, both had a lesson at different grade levels called `Fact or Fiction' or something along those lines. They were given pieces of information, a sentence, and then had to say whether it was true or nonsense. So, you can raise awareness in people that not everything you find on the Internet represents the truth.'' (Media Literacy and Media Education)}

\end{itemize}

\subsection{\final{Face-to-Face Conversations}}

\begin{itemize}

\item \final{CS-4: ``I think the opportunity lies in creating a discourse between people who know and like each other and are close. That is, between people who usually respect each other and each other's opinions. Thus the horizons can be broadened.'' (Conversation)}

\item \final{POLSCI-2: ``The most important technique remains direct conversations. But if we don’t have that, then possibly some browser add-ins that allow one to engage in dialogue.'' (Conversation, Socratic Method, and Inquiry)}

\item \final{POLSCI-2: ``And it seems to be --- and it seems plausible to me --- that it is better not to go directly into confrontation and ridicule everything, but instead to ask: Where does this information come from? Why do you believe this piece of information in this case, when in another case you would find it unlikely?' (Conversation, Socratic Method, and Inquiry)}

\end{itemize}

\subsection{\final{Crowdsourcing}}

\begin{itemize}

\item \final{CS-4: ``I could imagine a tool where you would invite people who know each other into the conversation. That is, if I believe that Biden stole the election and I read some news supporting that, for example, a notification could pop up on A.'s [a colleague] phone or laptop saying I’ve just viewed this, and they would have the opportunity to engage with me in a dialogue through this tool.'' (Social Network (People) / Social Proof)}

\item \final{CS-4: ``And then we have a kind of snowball system, where A. [a colleague] might think, `Yeah, that election thing is kind of weird.' And they rate it as trustworthy, but then it spreads to J. [another colleague], who then says, `I’m not so sure,' which then gets back to us, and we see that J. actually does not trust this headline much. `Why could that be?' And then we either have a platform where we could talk about it or something as simple as thumbs up or thumbs down [button].'' (Social Network (People) / Social Proof)}

\item \final{CS-4: ``In the end, people who are already convinced of such things can only be persuaded by people from their close circle or primarily by them, that it might not be true. This means some kind of social network where this can be discussed.'' (Social Network (People) / Social Proof)}

\end{itemize}

\subsection{\final{Labels For News Sources}}

\begin{itemize}

\item \final{PSY-3: ``Ideally, the system would store reliable sources and provide information from them. Of course, it requires significant upfront work to check: `What are reliable sources? And where can you accept the information, and where not?''' (Labeling of Reliable Sources)}

\item \final{LAW-4: ``One could try to do something with labeling, indicating that certain sources are more trustworthy than others.'' (Labeling of Reliable Sources)}

\item \final{COM-2: ``Certainly from a pool of verified sources, where it is known that these providers or news outlets are verified on the topic in question, and the quality is correspondingly high.'' (Labeling of Reliable Sources)}

\end{itemize}

\subsection{\final{Labels For News Articles}}

\begin{itemize}

\item \final{PSY-4: ``Even just labeling false statements or, as in the case of Donald Trump, at least marking that there is no evidence for it, is very helpful, I believe. But I honestly do not know of other systems like this.'' (Labeling of Disinformation)}

\item \final{BIO-1: ``If, for example, you click on something or read something, and it has a high likelihood of being disinformation, it would be great if it were tagged.'' (Labeling of Disinformation)}

\item \final{CS-1: ``[Computer Science] researchers [he is collaborating with] are working hard on this and have applied something called Nutrition Labels --- like what you see on food packaging --- to fake news.'' (Labeling of Disinformation)}

\end{itemize}

\subsection{\final{Finding Sources For Claims}}

\begin{itemize}

\item \final{LAW-2: ``In a scientific context, where claims are expected to be backed by evidence, you could demand that from an app. But in journalistic articles, where sources are often not cited, it is difficult to verify information [since sources are] not commonly provided.'' (Citing Sources)}

\item \final{POLSCI-4: ``The system could, for example, cite scientific studies or point to scientific findings, and thus provide a higher level of explanation or truth or, well, probability.'' (Citing Sources)}

\item \final{SOC-1: ``But the advantage would be if reports that are already published had sources that could be reconstructed and evaluated by users.'' (Citing Sources)}

\end{itemize}

\subsection{\final{Finding Information About Claims}}

\begin{itemize}

\item \final{POLSCI-3: ``We have clear numbers that show an excess mortality of around ten percent [due to COVID-19], which would be correct in this case. So, you would need to go to the source of the data to disprove the argument. In this case, it would be the Robert Koch Institute [a German federal government agency and research institute responsible for disease control and prevention], or the data on which the Robert Koch Institute bases its argument.'' (Disproving Disinformation)}

\item \final{CS-2: ``And if I see something where someone says masks are useless and we can’t breathe, it would be most interesting to see a study showing that this is not an issue. Or not just a study, but maybe the next YouTube video or something like that.'' (Citing Sources)}

\item \final{PSY-2: ``It would help me, at least, if I could trace what the actual data is, how it was interpreted, and why. It really always depends on the specific case.'' (Citing Sources)}

\end{itemize}

\subsection{\final{Finding Information About News Sources}}

\begin{itemize}

\item \final{LAW-2: ``People expect that well-known sites trying to mislead them are flagged. For example, you might say a source is questionable because a lot of disinformation is known from it.'' (Evaluating Sources)}

\item \final{CS-2: ``I sometimes wonder why we talk so much about big AI, when much can already be identified based on the sources alone, whether it is fake news or not. A lot could already be done there.'' (Evaluating Sources)}

\item \final{SOC-2: ``I guess I am old fashioned. I always sort of go by if I am reading something, I would see what is this source, right? You know, what references are cited are here and then I have an understanding. I might look them up.'' (Evaluating Sources)}

\end{itemize}

\subsection{\final{Supporting Source Comparison}}

\begin{itemize}

\item \final{POLSCI-3: ``So that I can compare disinformation with the verified sources or reliable information. And say, `Okay, what do these sources say, what do the others say, and where do these sources come from?'.''~(Comparing Sources)}

\item \final{LAW-1: ``Or, listen to what someone else has to say about it. By making comparisons, you can ultimately draw conclusions.''~(Comparing Sources)}

\item \final{CS-1: ``If I put myself in the user's shoes, I would want to upload a URL. This tool should scan the text, go through the images, and cross-reference them with other sources (...) from all over the world. So not just sources from, let’s say, Axel Springer, but from different sources, possibly even including social media.''~(Comparing Sources)}

\end{itemize}

\subsection{\final{Deleting Misinformation}}

\begin{itemize}

\item \final{LAW-4: ``Or like in social networks, where there are deletion requirements.'' (Deleting Disinformation)}

\item \final{POLSCI-4: ``I also think of those social media approaches, where things are deleted because it is clear the comment is wrong or contains something that is unacceptable from various perspectives --- racist, discriminatory, etc.'' (Deleting Disinformation)}

\item \final{COM-1: ``But I find it crazy that these pages could exist for so long and continue to operate. Some of it is still considered free speech, I know. But some of it is a mix of made-up stuff and things that are clearly meant to damage our political system. I think it is too easy to say Facebook should just step in. I do not get why we cannot discuss taking this kind of thing off the internet. It is not always that simple, but I am not saying this because I am a hardliner, only because I think there is a certain risk when we say we should make people aware of something that may take a decade or so, but right now it is causing a lot of harm.'' (Deleting Disinformation)}

\end{itemize}

\end{document}